# Journalists, media and influencers: An analysis of the conversation in the digital public sphere during the Qatar 2022 World Cup

**Simón Peña-Fernández[1]*, Ainara Larrondo-Ureta[2] and Jordi Morales-i-Gras[3]**

1. University of the Basque Country (UPV/EHU) (ROR: 000xsnr85)
   simon.pena@ehu.eus / https://orcid.org/0000-0003-2080-3241
2. University of the Basque Country (UPV/EHU) (ROR: 000xsnr85)
   ainara.larrondo@ehu.eus / https://orcid.org/0000-0003-3303-43307
3. University of the Basque Country (UPV/EHU) (ROR: 000xsnr85)
   jordi.morales@ehu.eus / https://orcid.org/0000-0003-4173-3609

* Corresponding author

**Abstract**: Public digital conversation around major sporting events takes place within a hybrid system in which journalists and the media compete with new intermediaries, including influencers, to gain greater visibility and engage with audiences. This study analyses the Qatar 2022 World Cup as a case of high informational intensity and public opinion monitoring. To that end, social network analysis was applied to X/Twitter using the hashtag #Qatar2022, analysing 1,343 high-engagement accounts—including those of journalists, media and influencers—alongside a random sample of 5,000 users. The findings indicate that journalists are under-represented in the user population as a whole, but significantly over-represented among the highest-engagement accounts, and they maintain stable visibility. The media, by contrast, attract a lower average level of attention and tend to achieve only sporadic peaks of impact. Accordingly, journalistic authority on social media is observed less as dominance in terms of participation volume and more as the capacity to occupy reference positions when public attention is being shaped during the event.

## 1. Introduction

The media have traditionally been attributed a central role in setting citizens' news agendas and shaping public opinion (McCombs, 2006). Within this conception of communication, relations between the media and other actors were markedly unidirectional and were characterised by the primacy of specific topics—most notably political information—and by a strong presence of official sources (Larrondo et al., 2024).

This is the author-accepted version of an article published in *Sapienza: International Journal of Interdisciplinary Studies*. The final version is available at: https://doi.org/10.51798/sijis.v7i2.1482

However, the transformation of the digital ecosystem into a hybrid media system (Chadwick, 2017) has eroded the role and centrality of traditional media (Pérez-Latre, 2014), which must now compete in an increasingly crowded environment with new interlocutors (Salaverría & Martínez-Costa, 2021). In the digital context, audience interaction has expanded beyond traditional media, with social media becoming a new space for the formation of public opinion (Masip et al., 2019).

This shift affects two classic foundations of journalism: control over the production and the distribution of information. Consequently, this new scenario invites analysis of the extent to which the media retain their capacity to generate news agendas and public debate on platforms and social media, where their visibility competes with that of other actors who also actively seek interaction (van Dijk et al., 2018). All of this unfolds in a space in which traditional editorial logics coexist with platform logics, and where competition is not confined to information production but extends to the ability to capture audience attention in a context of information abundance (Napoli, 2011).

In sport, social media have offered an excellent opportunity for clubs and athletes to find their own voice and connect directly with followers, without intermediaries (Pegoraro & Jinnah, 2012; Cano & Paniagua, 2017; Fenton et al., 2021). As a result, they have been able to build high-impact brands that allow them to drive followers to their official websites, generate interaction, and strengthen both a sense of community belonging and commercial revenues (Meso-Ayerdi et al., 2021).

Yet, although this change enables an unprecedented intensity in relationships with fans—as reflected in continuous access to players' lives or instant match updates—it also raises challenges such as athletes' overexposure to public opinion and potential cognitive fatigue (Sanderson, 2011). The intermediary role traditionally played by the media is also affected (Price et al., 2013), while the need to produce large volumes of content at speed, driven by social media dynamics, may condition journalism's functions of contextualisation and verification.

For journalists, direct contact between sources and audiences has likewise entailed a loss of centrality and prominence in an environment where the boundaries between traditional communication roles are increasingly blurred, and in which the definition of what it means to be a journalist is becoming increasingly liquid (Deuze & Witschge, 2017). Traditional principles of authority and legitimacy—that is, the capacity to exert some form of social influence—have extended into social media (Casero-Ripollés, 2020), which have begun to emerge as a "fifth power" (Pérez-Escoda & Rubio-Romero, 2022). Journalists therefore increasingly share the capacity to generate information and opinion with other authoritative figures, such as audiences and new intermediaries (Turner, 2015). This has led them to adopt an ambivalent stance towards social media use, simultaneously seeking to align themselves with and differentiate themselves from the practices of audiences and other information actors (Price et al., 2013; Tworek & Buschow, 2016). Audience engagement is perceived as the means to achieve a more enduring and mutually beneficial relationship with audiences (Belair-Gagnon et al., 2018)

On platforms and social media, the new sources of authority are not the institutional or social functions fulfilled by different actors, but rather their ability to generate views, interactions, and occupy space in the digital public sphere. In any case, social media have reinforced a feature already present in modern societies, since mediatization is inseparable from major sporting events and constitutes the only means by which millions of followers can experience and share events that only a very limited number of people can attend live at the venue (Rowe, 2012). Sport is a global phenomenon that transcends linguistic barriers and crosses national borders (Miller et al., 2001). In such events, debate and dialogue among users are also intrinsic to the event itself, and active communities are formed around athletes and teams, where elements of identity, affinity and interest overlap. Thus, despite sport being one of the domains in which globalisation manifests itself most clearly, the fact that global spectacles such as the World Cup are organised around sporting nations means that these continue to play a fundamental role (Rowe, 2012).

These characteristics are particularly relevant in the case of hegemonic sports such as football, whose activity extends beyond the sporting realm and has become one of the main expressions of popular culture (Miller et al., 2001). In football, the struggle for referential authority in the digital public sphere is intensified by its massive and cross-cutting following across multiple places and cultures, which has in turn transformed major clubs into multinational entertainment corporations (Ginesta, 2011).

Previous research suggests that network structures on X/Twitter allow the prominence of major sporting organisations to be further recognised and legitimised, while individual citizens have a limited capacity to generate influence. In this regard, the FIFA World Cup held in Qatar offered an excellent opportunity to analyse how debates develop in the digital public sphere, given its enormous capacity to capture audience attention. According to data provided by FIFA (2024), the final between France and Argentina was watched by 1.42 billion people, while average viewership per match in the final stage reached 175 million worldwide.

From a communication perspective, this sporting event was simultaneously an opportunity to promote the host country’s image in the international press through “football diplomacy” (Bianco & Sons, 2023; Næss, 2023; Dubinsky, 2024), and a source of controversy due to alleged sportswashing linked to the numerous issues surrounding its organisation (Brannagan & Reiche, 2022; Samuel-Azran et al., 2022; Grix et al., 2023; Jain, 2023).

As a result, debate on social media about the Qatar 2022 World Cup revolved for an extended period around political and cultural issues beyond strictly sporting matters (Hassan & Wang, 2023), and existing studies indicate that the host country’s image on social media was not predominantly positive (Dun et al., 2022). Research to date has focused on the influence of sporting organisations, celebrities and public figures (Yan et al., 2018; Hassan & Wang, 2023; Zákravský et al, 2024), but the presence of journalists and media in conversations within the digital public sphere during high-impact sporting events has not been examined.

Against this backdrop, the present study aims to examine the role of journalists and the media as reference actors in the digital public sphere through a high-impact sporting event such as the Qatar 2022 World Cup. To this end, the following research questions are proposed:

RQ1. What has been the chronological evolution of social media conversation about the Qatar 2022 World Cup?

RQ2. Do journalists and media organisations exhibit greater referential authority and engagement capacity in this global conversation than other users?

RQ3. Is the referential authority of journalists and media organisations in the global conversation a consolidated phenomenon?

## 2. Method

To address the proposed research questions, the study employed Social Network Analysis (SNA) on X/Twitter. On the one hand, X is the platform most commonly used by public figures to express their views; on the other, SNA makes it possible to analyse relationships between users and their interactions through likes, replies, reposts (retweets), quotes and mentions. This approach not only allows the diffusion of messages to be quantified, but also enables the identification of interaction patterns and the centrality positions occupied by senders in relation to a given event.

Access to X data was obtained via the Academic Research track of the API 2.0, which is no longer available. The hashtag used for data collection was #Qatar2022. Prior to sample selection, an analysis was conducted to determine the most frequently used hashtags related to the tournament (Figure 1). Among the ten most widely used hashtags, #Qatar2022 was by a wide margin the most popular option, having been used on more than one million occasions. To a lesser extent, other hashtags referring to the tournament and to national teams—particularly Argentina—were also observed.

Figure 1. Frequency of hashtag use

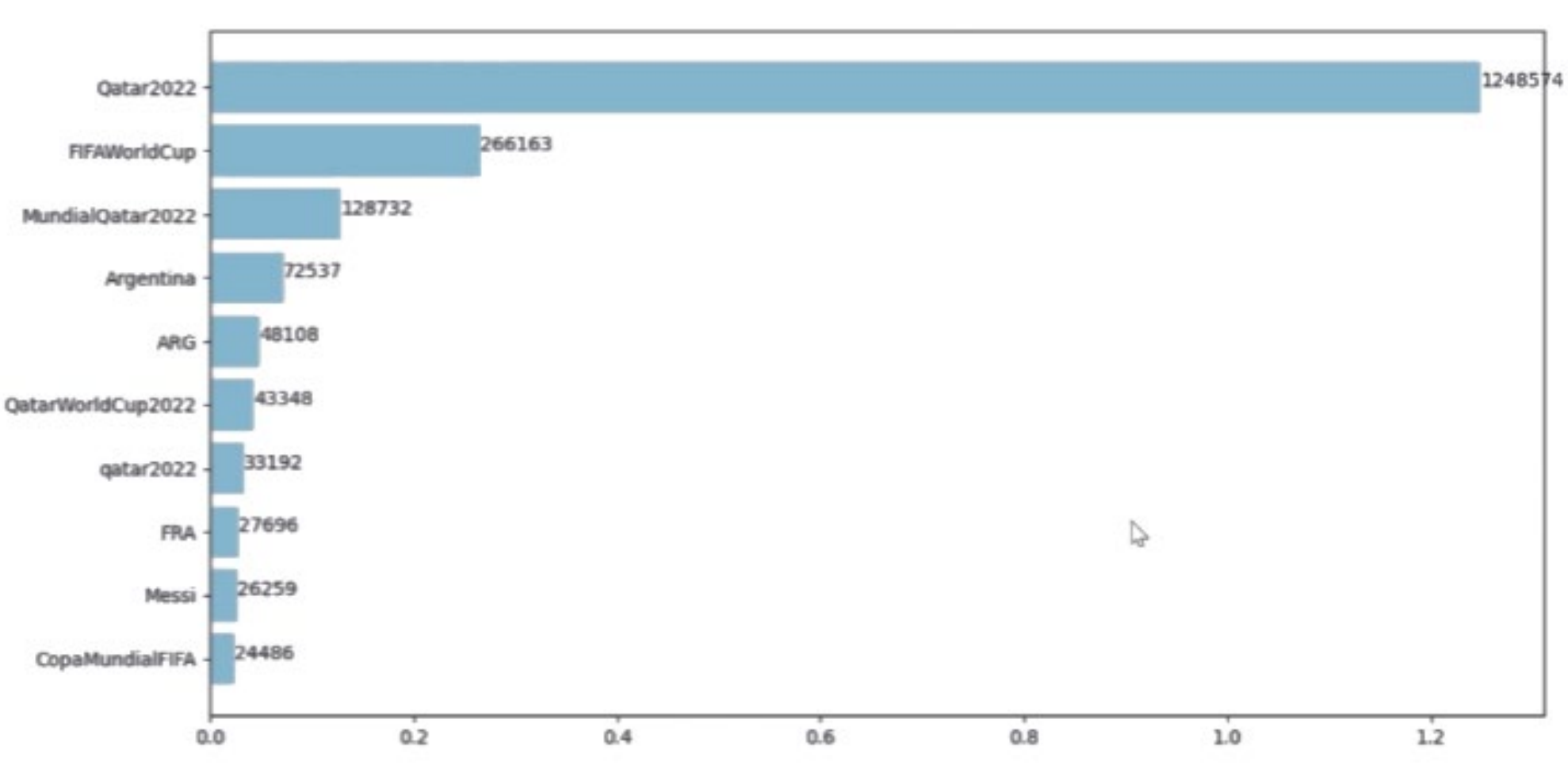

Source: Authors' own elaboration

To analyse the chronological evolution of the conversation about the World Cup (RQ1), the number of messages published with the hashtag #Qatar2022 was measured over the period from 1 January to 31 December 2022. This temporal design made it possible to capture both the pre-event phase (qualification, draw, controversies surrounding the organisation, etc.), the period of maximum intensity (the final stage), and the stage immediately following the final. The final stage of the tournament took place in Qatar between 20 November and 18 December.

To establish the typology of users participating in the conversation, the engagement achieved by their messages and the standard deviation of that engagement were examined (RQ2). In other words, the analysis measured the success of the published messages and whether this success was achieved in a sustained or sporadic manner (RQ3). Engagement was operationalised as the sum of reposts (with and without comment), likes and replies. For each user, both the average engagement achieved and its standard deviation were calculated.

A total of 1,343 high-engagement users were identified, whose data were analysed using an LLM based on Llama 3.1. The sample was complemented with a random sample of 5,000 general users. Based on this analysis, the following user categories were generated:

Table 1. Number of high-engagement users, by type

| Individual accounts | | Collective / institutional accounts | |
|---|---|---|---|
| Academic | 189 | Company | 149 |
| Activist | 52 | Government/Politics | 43 |
| Artist | 408 | Media organisation | 617 |
| Athlete | 142 | Sports organisation | 245 |
| Entrepreneur | 121 | Other | 250 |
| Actor/Actress | 31 | | |
| Journalist | 1,228 | | |
| Politician | 58 | | |
| Professional | 290 | | |
| Public figure | 34 | | |
| Influencer | 71 | | |
| Student | 23 | | |
| Other | 1,049 | | |

Source: Authors' own elaboration

The analysis of the relationship between generated engagement and its standard deviation made it possible to identify four types of users during the period analysed: common users, home runners, influencers and rising stars.

First, the largest number of users typically achieved low engagement with their messages. This category of common users would encompass the vast majority of the accounts analysed.

The home runners category, for its part, includes users who on average tend to obtain low engagement, but who during this period saw one of their messages achieve a very high level of interaction, resulting in a high standard deviation. To assign this category, users with an average engagement below 100 and a standard deviation above 50 were considered. This category comprised 808 users.

Third, influencers are those who regularly achieve high engagement, with little standard deviation. To assign this category, users with an average engagement above 100 and a standard deviation below 50 were considered. In this conversation, five influential users with these characteristics were identified.

Finally, positioned between the previous two categories, is the group termed rising stars: those who achieved substantial engagement, albeit with a significant standard deviation. To assign this category, users with an average engagement above 100 and a standard deviation above 50 were grouped. This category comprised 530 users.

In all cases, only users with at least six posts during the analysis period were considered, in order to avoid outliers and extreme values.

## 3. Results

*3.1. Evolution of interest*

The chronological analysis of the hashtag #Qatar2022 (Figure 2) clearly shows the key newsworthy moments of the event. The point of greatest attention on social media was the final between Argentina and France, played on 18 December 2022. This milestone crowned a series of moments of heightened attention in the digital public sphere that had begun a month earlier with the opening match of the World Cup's final stage between Qatar and Ecuador.

Figure 2. Evolution of the hashtag #Qatar2022

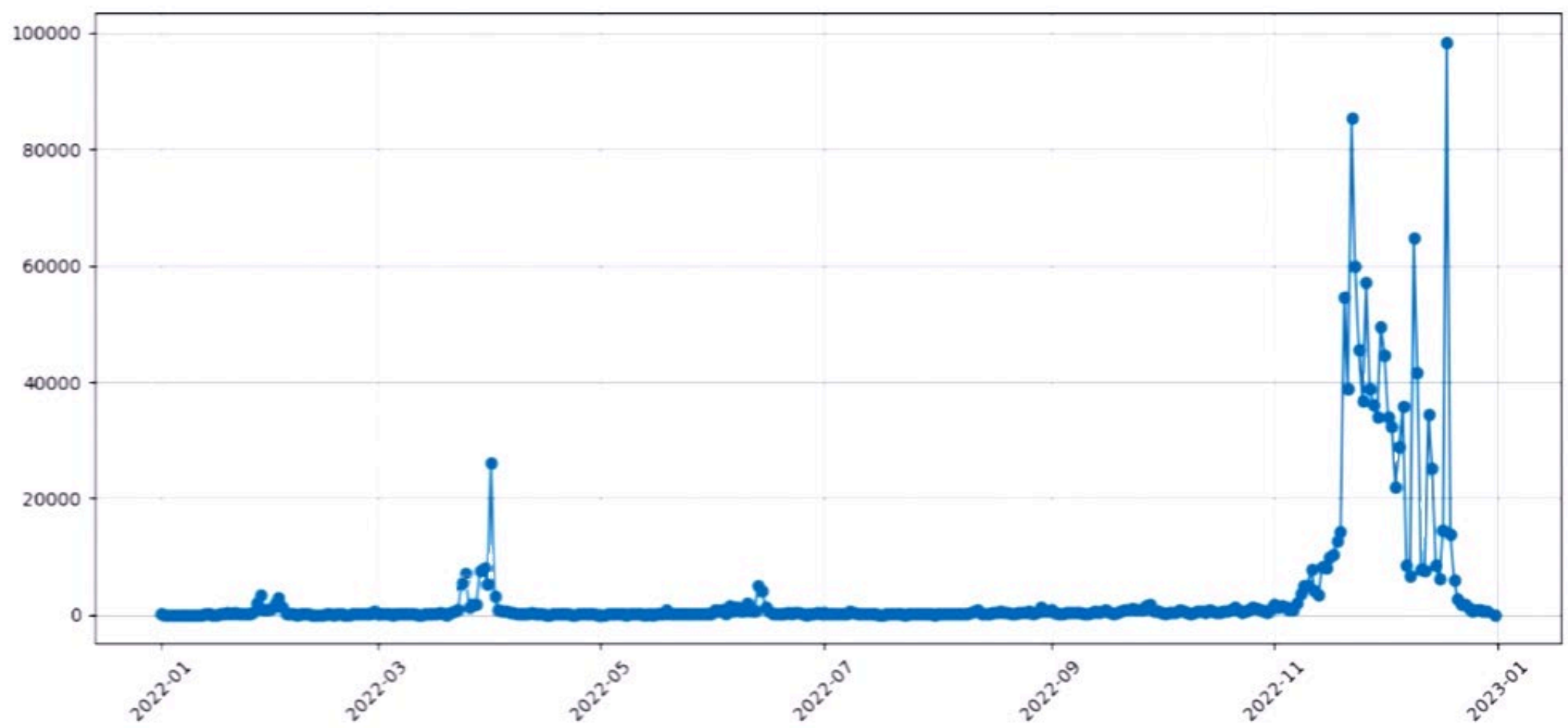

Source: Authors' own elaboration

Earlier, audience attention had also been drawn—albeit to a much lesser extent—by the last decisive matches of the qualification phase across the different continents (between 24 and 29 March 2022) and, in particular, by the group-stage draw for the final tournament, held on 1 April 2022.

Figure 3. Types of users by generated engagement

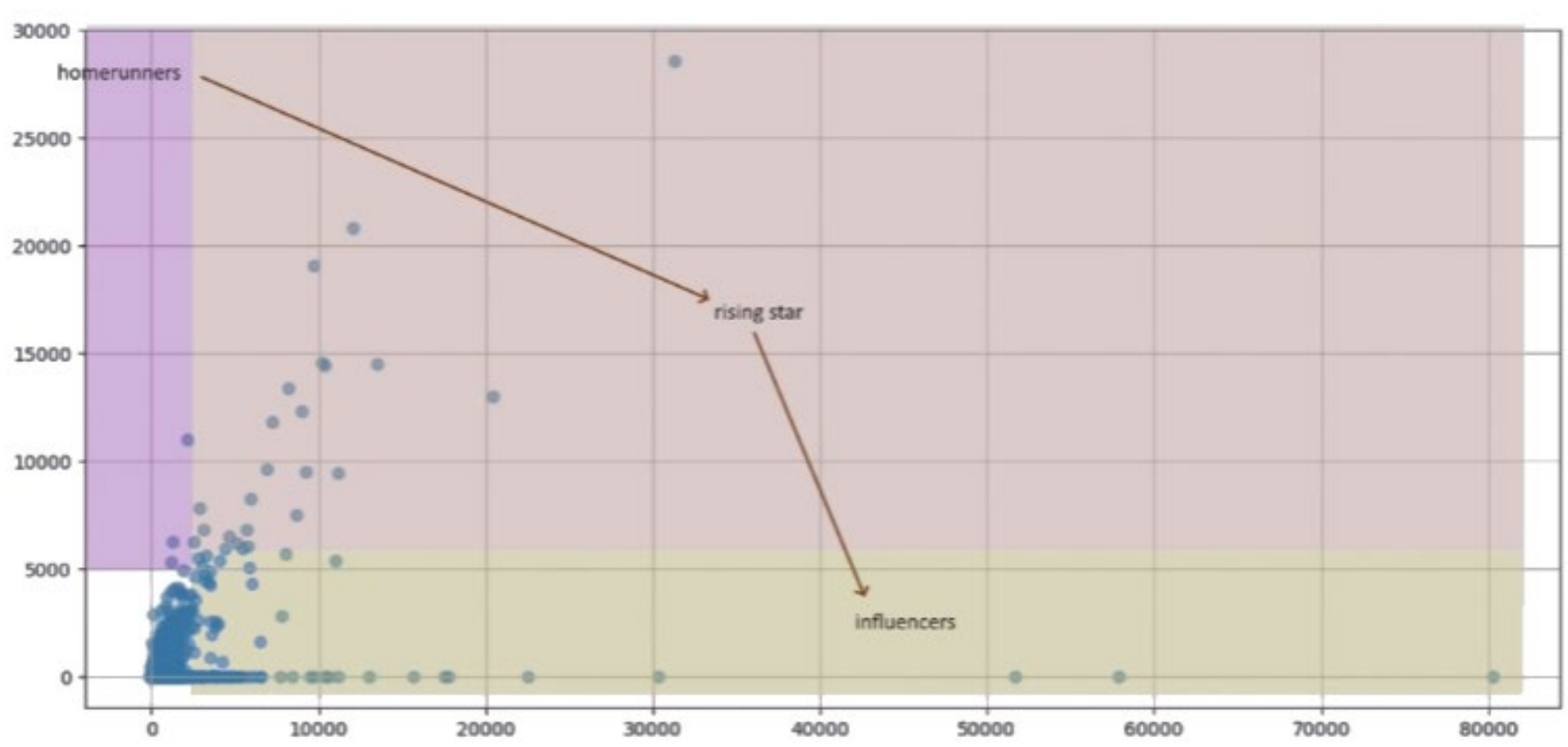

Source: Authors' own elaboration

The analysis of this activity suggests a conversation driven by specific milestones and lacking continuity, reinforcing the idea that the World Cup functioned as a media event that concentrated attention abruptly and redistributed it towards other topics with equal speed. In terms of competition among different actors for audience attention in the digital public sphere, this temporal concentration is significant because it points to a key constraint in the struggle for visibility: those who already enjoy prior recognition are able to consolidate their presence on social media through events of this kind, whereas the emergence of new reference figures depends on brief and highly competitive windows.

The vast majority of accounts were located in the low-engagement space (Figure 3). Among high-engagement users, home runners—that is, those who achieved high levels of attention with only some of their messages—accounted for 60.3% of the analysed sample. Users classified as rising stars, who attracted substantial audience attention with several of their messages but with fluctuations over time, represented 39.5% of the analysed users. Finally, influencers, understood as those users who consistently achieved high levels of attention in the digital public sphere across all their messages, accounted for only 0.4% of the sample. This imbalance confirms the existence of a strong concentration of attention on platforms in the hands of a very small number of users.

*3.2. Referential authority and engagement capacity*

In the analysis of the data by user type, as shown in Figure 4, journalists stand out for their divergent behaviour across the two groups examined: users overall and users who achieve high levels of influence or engagement.

Figure 4. Engagement by type of individual account

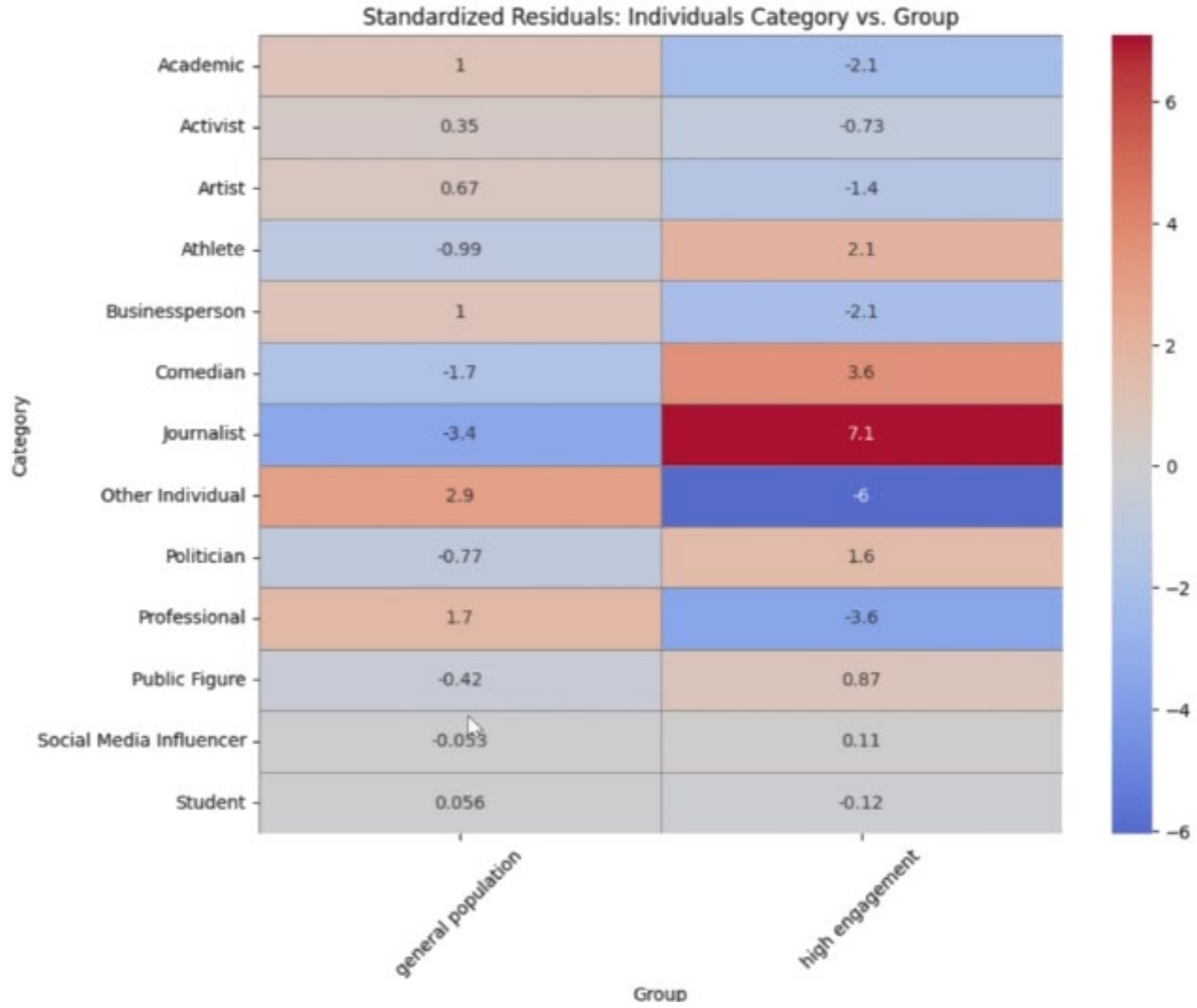

Source: Authors' own elaboration

When analysed within the global set of users, the value is −3.4, indicating that there are fewer journalists than expected in this group. Residuals greater than ±2 are considered statistically relevant; therefore, this represents a significant difference, showing that journalists are under-represented in this group compared with what would be statistically expected.

By contrast, within the group of the most influential users—those who achieve the highest levels of engagement—the value is +7.1, an extremely high figure that indicates a very significant over-representation of journalists. This suggests that journalists not only participate actively, but also occupy central positions in spaces of high visibility within the global conversation.

This finding is also consistent with journalists' long-standing presence on X/Twitter, but the results point to an important distinction: presence does not equate to centrality. Journalists do not dominate the overall user population, yet they feature prominently in the spaces where interaction is most concentrated. This over-representation suggests that their professional capital and their position in the digital public sphere remain relevant on platforms, even though they are now expressed through metrics of visibility and engagement.

Figure 5. Engagement by type of collective account

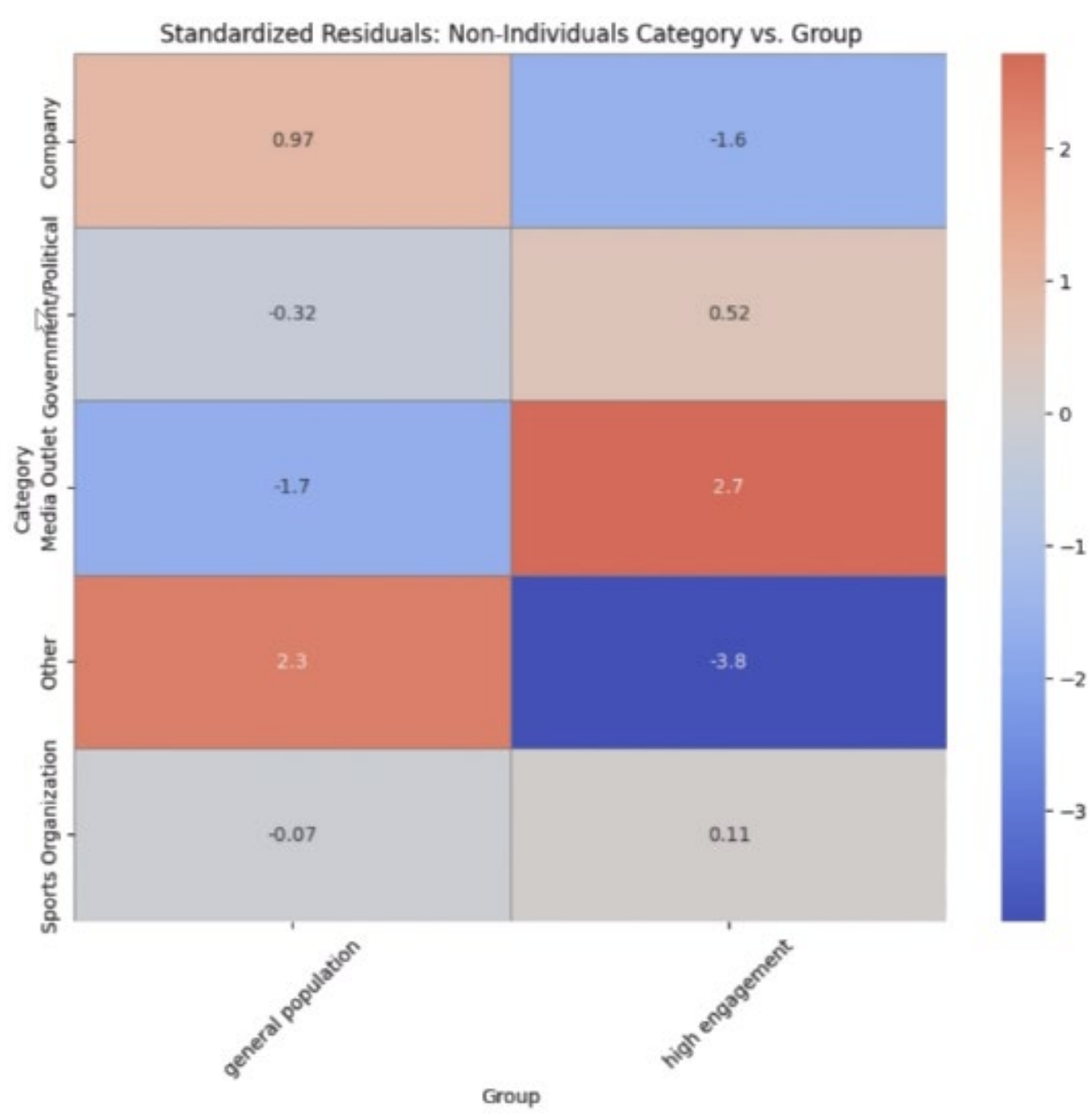

Source: Authors' own elaboration

In the case of news media, the table of standardised residuals for collective or institutional accounts (Figure 5) yields results similar to those observed for journalists, although with less extreme values. When compared with the overall set of non-individual accounts, the value is −1.7, indicating that these entities are less present than would be expected if there were no relationship between category and level of participation. Although this does not reach the conventional threshold of statistical significance (±2), it comes close and already suggests a tendency towards under-representation.

However, within the group of accounts with high referential authority and engagement, the residual is +2.7, meaning that there are notably more news media accounts than expected. This value exceeds the ±2 threshold, allowing us to state that this over-representation is statistically significant. As with journalists, media outlets play a prominent role among the collective accounts that generate the highest levels of interaction.

These findings also suggest that individual accounts tend to achieve higher engagement than collective accounts. The publication of content framed in a more generic and neutral manner by institutional accounts may be one of the reasons for this lower level of interest and for the dilution of average interaction per message. Even so, the over-representation of media outlets among high-engagement accounts indicates that they retain the capacity to activate audience attention when intervening at sensitive points in the news cycle.

*3.3. Consistency and effervescence*

Finally, the analysis focused on those accounts that achieved the highest levels of engagement and on the frequency or consistency with which this was attained (Figure 6), defining home runners as those who achieved a high impact only very occasionally, and rising stars as those who did so more regularly. Users classified as influencers were not considered, as this group was small and not statistically significant. Among home runners, the residual is −0.36, indicating that there are slightly fewer journalists than expected, although the value is very low and not statistically significant. Among rising stars, the residual is +0.45, suggesting a slight over-representation, again within normal statistical margins.

Accordingly, unlike other categories (such as athletes), journalists do not stand out particularly in either of these two groups. This may be because journalists already enjoy a certain level of structural and stable visibility, and therefore do not tend to be among the users whose visibility increased during the World Cup. In other words, they maintained a constant, but not explosive, presence.

Figure 6. High-engagement users, by type of individual account

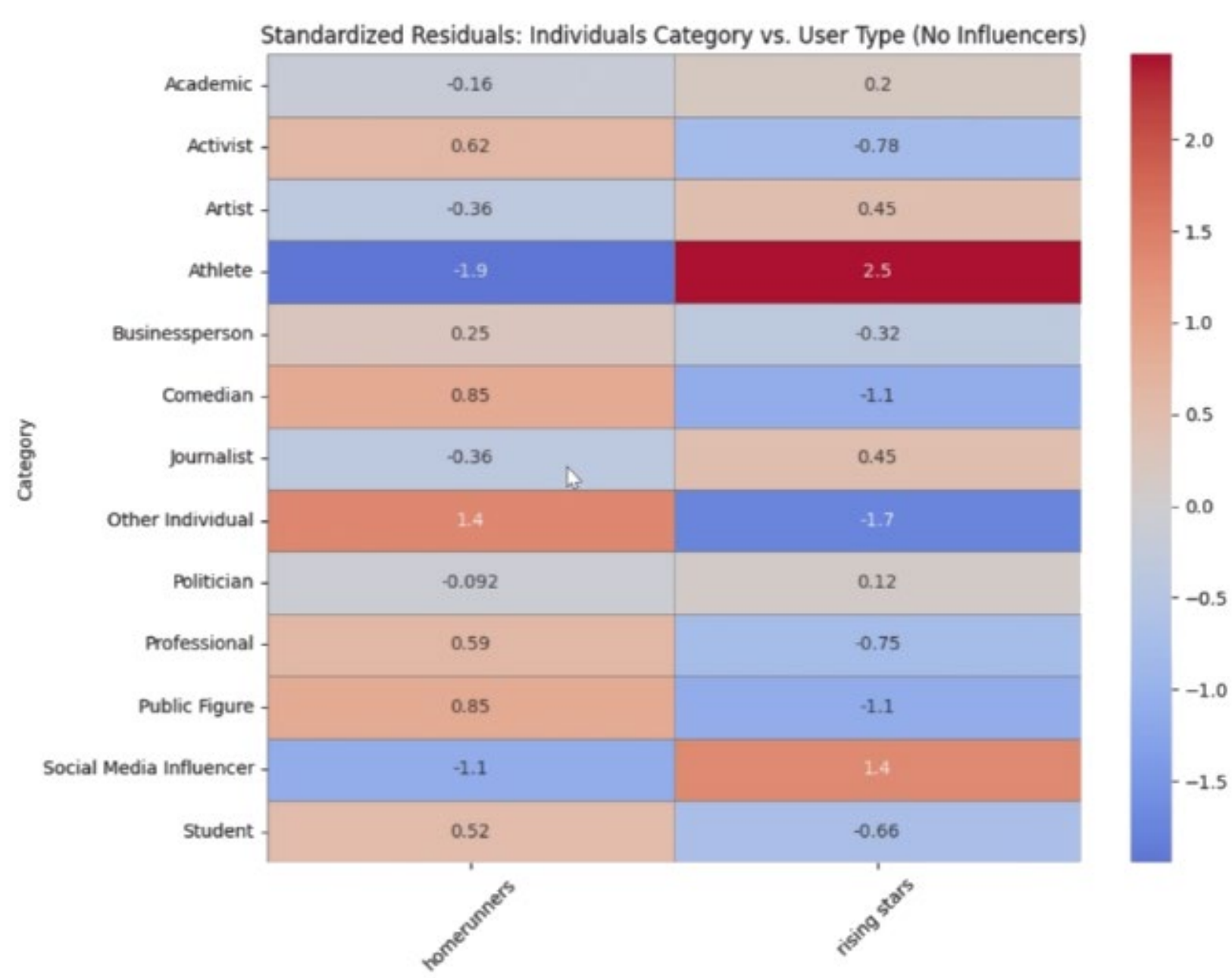

Source: Authors' own elaboration

These data are also consistent with the attention-related benefits of maintaining a stable presence on social media, which means that the visibility of journalists and media outlets depends less on occasional breakout moments and more on recognition derived from a consistent prior trajectory. In other words, high-news-value events such as the World Cup do not create influential journalists or media organisations; rather, they provide a stage on which users with pre-existing relational capital can capitalise, while growth occurs primarily among profiles linked to the sporting field and specifically associated with the nature of the event.

By contrast, in the case of media organisations (Figure 7), the data reveal the opposite pattern. Within the home runners category—that is, those accounts that occasionally exceed their expected levels of engagement—the residual is +2.1, indicating that there are more entities of this type than expected. Conversely, the presence of media outlets in the rising stars group is significantly lower than expected (−2.5).

This combination suggests that news media enjoy a consolidated and habitual presence, but that, given the high volume of messages they publish and the neutral tone typical of institutional accounts, they tend to achieve lower average levels of attention. In other words, they produce content far more consistently than individual accounts, but with a more limited capacity to generate engagement. For their part, growth profiles among collective accounts are, as might be expected given the nature of the event, mainly found among sporting institutions.

Figure 7. High-engagement users, by type of collective account

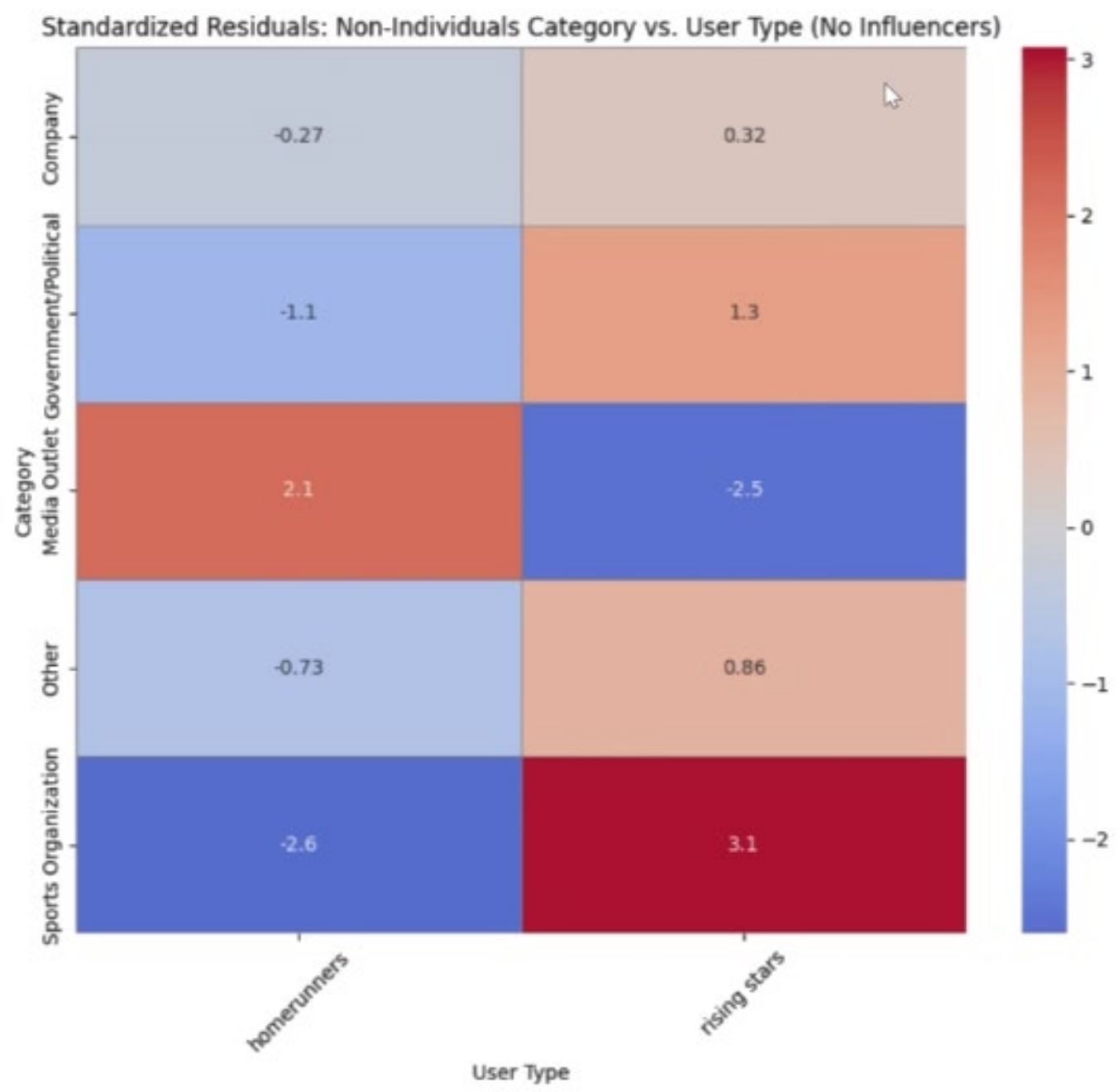

Source: Authors’ own elaboration

## 4. Conclusions

This study demonstrates that, in the context of the Qatar 2022 FIFA World Cup, social media—particularly X/Twitter—became consolidated as a key space for public conversation, in which journalists and news media maintained a central role, albeit one increasingly shared with new actors such as influencers and highly communicative individual users.

The findings reveal a statistically significant over-representation of journalists and media organisations among the groups that achieved the highest levels of engagement and influence. This reinforces the idea that traditional media authority persists in digital environments, albeit adapted to the new logics of platforms. This greater centrality does not mean that journalists and media capture a substantial share of the overall conversation; rather, it manifests in their over-representation within spaces of high interaction. In other words, journalists and media do not dominate the global conversation, but they retain the capacity to occupy reference positions when attention becomes concentrated around a major news event.

However, this authority is not expressed uniformly. Journalists tend to maintain structural and stable visibility, whereas institutional media accounts display a more episodic capacity for impact. This reflects a dynamic of content saturation and constant communicative competition. The distinction between journalists and media outlets is relevant, as it suggests that personalisation—and the recognition associated with individual voices—remains an asset on X/Twitter, while the corporate logic of high publication volume combined with more neutral content tends to reduce average attention.

Moreover, the user typology employed provides a more nuanced understanding of how attention is constructed on social media. The typology proposed here helps to disentangle phenomena that are often conflated: (1) visibility as a one-off occurrence (home runners), (2) sustained attention (influencers), and (3) growth during the event (rising stars). In a high-intensity media event such as the football World Cup, this distinction is particularly relevant, as conversation is structured around brief windows of attention and a distribution of messages that rewards temporal opportunity.

From the perspective of analysing the information ecosystem, these findings support the notion of a hybrid media system, in which institutional logics of authority coexist with emerging dynamics based on the ability to generate audience attention. This poses strategic challenges for media organisations, such as balancing editorial principles with the need to attract audience attention in a context of highly fragmented audiences in which they no longer occupy a central position. In other words, the hybridisation of the information ecosystem does not eliminate journalistic authority, but it does displace it from a regime of structural authority—where it occupied the centre of the communicative space—to one in which legitimacy is also measured in terms of audience attention and where its position is more peripheral.

From an applied perspective, media organisations and journalists seeking to strengthen their leadership on social media should consider not only publication frequency, but also message personalisation, dialogic quality, the strategic use of hashtags, and adaptation to the real-time dynamics of global media events. The issue, therefore, is not merely being present on social media and publishing frequently, but how this is done in an environment where users directly address audiences. In high-intensity information contexts, constant publication may sustain presence, but it does not guarantee sustained interaction, especially for institutional accounts. By contrast, the articulation of a recognisable voice tends to favour the occupation of reference positions.

Finally, the analysis of the global conversation during the football World Cup held in Qatar, and particularly the presence of journalists and media organisations within it, confirms that major global sporting events generate massive conversation in which audience attention is not distributed horizontally. Actors with pre-existing capital (journalists, media, influencers) are better positioned to capitalise on the capacity to generate attention. This does not contradict the openness of the digital space, but it does invite us to interpret it as a competitive arena, where visibility is a resource and influence is achieved under conditions of structural inequality. The position of journalists and media is more peripheral in the global conversation, yet they stand out for their capacity to become visible among the most influential users.

This study has several limitations. First, the analysis is based on the hashtag #Qatar2022, which helps delimit the conversation, reduce noise, and allows for an in-depth examination of discourse around that tag. Nevertheless, part of the debate surrounding the World Cup may have taken place under other hashtags or variants—for example, depending on linguistic communities. In addition, the study focuses on X/Twitter, which does not represent public opinion as a whole, not even within the digital sphere. However, it does constitute a useful space for observing how different actors compete for visibility and seek to become reference points during events of high news visibility.

## References

Bélair-Gagnon, V., Nelson, J.L., & Lewis, S.C. (2018). Audience Engagement, Reciprocity, and the Pursuit of Community Connectedness in Public Media Journalism. Journalism Practice, 13, 558-575. https://doi-org.ehu.idm.oclc.org/10.1080/17512786.2018.1542975

Bianco, C., & Sons, S. (2023). More than a Game: Football and Soft Power in the Gulf. The International Spectator, 58(2), 92–106. https://doi.org/10.1080/03932729.2023.2196810

Brannagan, P. M., & Reiche, D. (2022). Qatar and the 2022 FIFA World Cup Politics, Controversy, Change. Palgrave Macmillan.

Cano Tenorio, R., & Paniagua Rojano, F. J. (2017). El uso de Twitter por parte de los futbolistas profesionales. Contenidos y relaciones con los públicos. Revista Internacional De Relaciones Públicas, 7(13), 101–122. https://doi.org/10.5783/revrrpp.v7i13.457

Casero-Ripollés, A. (2020). Influence of media on the political conversation on Twitter: Activity, popularity, and authority in the digital debate in Spain". Icono 14, 18(1), 33-57. https://doi.org/10.7195/ri14.v18i1.1527

Chadwick, A. (2017). The Hybrid Media System: Politics and Power. Oxford University Press. https://doi.org/10.1093/oso/9780190696726.001.0001

Deuze, M., & Witschge, T. (2017). Beyond journalism: Theorizing the transformation of journalism. Journalism, 19(2), 165-181. https://doi-org.ehu.idm.oclc.org/10.1177/1464884916688550

Dubinsky, Y. (2024). Clashes of cultures at the FIFA World Cup: Reflections on soft power, nation building, and sportswashing in Qatar 2022. Place Branding and Public Diplomacy, 20, 218–231. https://doi.org/10.1057/s41254-023-00311-8

Dun, S., Rachdi, H., Memon, S. A., Pillai, R. K., Mejova, Y., & Weber, I. (2022). Perceptions of FIFA Men's World Cup 2022 Host Nation Qatar in the Twittersphere. International Journal of Sport Communication, 15(3), 197-206. https://doi.org/10.1123/ijsc.2022-0041

Fenton, A., Keegan, B. J., & Parry, K. D. (2021). Understanding Sporting Social Media Brand Communities, Place and Social Capital: A Netnography of Football Fans. Communication & Sport, 11(2), 313-333. https://doi-org.ehu.idm.oclc.org/10.1177/2167479520986149

FIFA (2024, 29 de noviembre). Dos informes detallan las cifras récord de audiencia global y los buenos resultados en sostenibilidad de la Copa Mundial de la FIFA Catar 2022™. InsideFIFA. https://inside.fifa.com/es/organisation/news/informes-cifras-record-audiencia-global-buenos-resultados-sostenibilidad-mundial-catar

Ginesta, X. (2011). El fútbol y el negocio del entretenimiento global. Los clubes como multinacionales del ocio. Comunicación y Sociedad, 24(1), 141-166. https://doi.org/10.15581/003.24.36228

Grix, J., Dinsmore, A., & Brannagan, P. M. (2023). Unpacking the politics of 'sportswashing': It takes two to tango. Politics, 45(3), 377-398. https://doi.org/10.1177/02633957231207387

Hassan, A. A. M., & Wang, J. (2023). The Qatar World Cup and Twitter sentiment: Unraveling the interplay of soft power, public opinion, and media scrutiny. International Review for the Sociology of Sport, 59(5), 679-704. https://doi.org/10.1177/10126902231218700

Jain, S. (2023). Resistance and Reform as Responses to Human Rights Criticism: Relativism at FIFA World Cup Qatar 2022. German Law Journal, 24(9), 1691–1702. http://doi.org/10.1017/glj.2023.119

Larrondo-Ureta, A., Peña-Fernández, S., & Morales-i-Gras, J. (2024). Política vs. entretenimiento. Análisis del liderazgo conversacional de los medios en las redes sociales. Profesional de la información, 33(4). https://doi.org/10.3145/epi.2024.0425

Masip, P., Ruiz-Caballero, C., & Suau, J. (2019). Audiencias activas y discusión social en la esfera pública digital. Artículo de revisión. Profesional de la información, 28(2), e280204. https://doi.org/10.3145/epi.2019.mar.04

McCombs, Maxwell. (2006). Estableciendo la agenda. El impacto de los medios en la opinión pública y el conocimiento. Barcelona: Paidós.

Meso-Ayerdi, K., Pérez-Dasilva, J.Á, & Mendiguren-Galdospin, T. (2021). Futbolistas en Twitter. Una plataforma para la autopromoción. Revista de Comunicación, 20(2), 277-301. https://doi.org/10.26441/RC20.2-2021-A15

Miller, T., Lawrence, G., McKay, J., & Rowe, D. (2001). Globalization and sport: Playing the world. SAGE. https://doi.org/10.4135/9781446218396

Napoli, P. M. (2011). Audience Evolution: New Technologies and the Transformation of Media Audiences. Columbia University Press.

Næss, H. E. (2023). A figurational approach to soft power and sport events. The case of the FIFA World Cup Qatar 2022™. Frontiers in Sports and Active Living, 5, 1142878. http://doi.org/10.3389/fspor.2023.1142878

Pegoraro, A.,& Jinnah, N. (2012). Tweet 'em and reap 'em: The impact of professional athletes' use of Twitter on current and potential sponsorship opportunities. Journal of Brand Strategy, 1(1). https://doi.org/10.69554/FWFD9361

Pérez-Escoda, A., & Rubio-Romero, J. (2022). Redes sociales, ¿el quinto poder? Una aproximación por ámbitos al fenómeno que ha transformado la comunicación pública y privada. Valencia: Tirant Lo Blanch.

Pérez-Latre, F. J. (2014). Legacy Media: A Case for Creative Destruction? Palabra Clave, 17(4), 1097-1113. https://doi.org/10.5294/pacla.2014.17.4.5

Price, J., Farrington, N., & Hall, L. (2013). Changing the game? The impact of Twitter on relationships between football clubs, supporters and the sports media. Soccer & Society, 14(4), 446–461. https://doi.org/10.1080/14660970.2013.810431

Rowe, D. (2012). Reflections on Communication and Sport: On Nation and Globalization: On Nation and Globalization. Communication & Sport, 1(1-2), 18-29. https://doi.org/10.1177/2167479512467328

Salaverría-Aliaga, R., & Martínez-Costa, M. P. (2021). Medios nativos digitales en España. Caracterización y tendencias. Sevilla: Comunicación Social. https://www.comunicacionsocial.es/libro/medios-nativos-digitales-en-espana_133545

Samuel-Azran, T., Hayat, T., & Galily, Y. (2022). Global Sport Protest Activism Is Exclusive to the Global Elite: A Case Study of #boycottqatar2022. American Behavioral Scientist, 67(10), 1179-1193. https://doi.org/10.1177/00027642221118299

Sanderson, J. (2011). How social media is changing sports: It’s a whole new ballgame. Hampton Press.

Turner, G. (2015). Re-inventing the media. London: Routledge. https://doi.org/10.4324/9781315675206

Tworek, H. J. S., & Buschow, C. (2016). Changing the Rules of the Game: Strategic Institutionalization and Legacy Companies' Resistance to New Media. International Journal of Communication, 10, 2119-2139. https://ijoc.org/index.php/ijoc/article/view/5179

van Dijck, J., Poell, T., & de Waal, M. (2018). The Platform Society: Public Values in a Connective World. Oxford University Press.

Yan, G., Watanabe, N. M., Shapiro, S. L., Naraine, M. L., & Hull, K. (2018). Unfolding the Twitter scene of the 2017 UEFA Champions League Final: social media networks and power dynamics. European Sport Management Quarterly, 19(4), 419–436. https://doi-org.ehu.idm.oclc.org/10.1080/16184742.2018.1517272

Zákravský, J., Gutiérrez-Chico, F., & Pulleiro Méndez, C. (2024). The political stance of Ibero-American national teams on social media towards Qatar during the 2022 FIFA World Cup. International Review for the Sociology of Sport, 60(3), 488-506. https://doi-org.ehu.idm.oclc.org/10.1177/10126902241274059Ananny, M., & Karr, J. (2025). How media unions stabilize technological hype: Tracing organized journalism’s discursive constructions of generative artificial intelligence. Digital Journalism, 1–21. https://doi.org/10.1080/21670811.2025.2454516.